# Contrastive and Transfer Learning for Effective Audio Fingerprinting through a Real-World Evaluation Protocol

Christos Nikou [1]✉, Theodoros Giannakopoulos[2]✉

[1,2]Multimedia Analysis Group of the Computational Intelligence Laboratory (MagCIL), Institute of Informatics and Telecommunications, NCSR "DEMOKRITOS", Athens, Greece.
[1] ✉ Corresponding author: chrisnick@iit.demokritos.gr
[2] ✉ Corresponding author: tyianak@iit.demokritos.gr

**ABSTRACT**

Recent advances in song identification leverage deep neural networks to learn compact audio fingerprints directly from raw waveforms. While these methods perform well under controlled conditions, their accuracy drops significantly in real-world scenarios where the audio is captured via mobile devices in noisy environments. In this paper, we introduce a novel evaluation protocol designed to better reflect such real-world conditions. We generate three recordings of the same audio, each with increasing levels of noise, captured using a mobile device's microphone. Our results reveal a substantial performance drop for two state-of-the-art CNN-based models under this protocol, compared to previously reported benchmarks. Additionally, we highlight the critical role of the augmentation pipeline during training with contrastive loss. By introducing low pass and high pass filters in the augmentation pipeline we significantly increase the performance of both systems in our proposed evaluation. Furthermore, we develop a transformer-based model with a tailored projection module and demonstrate that transferring knowledge from a semantically relevant domain yields a more robust solution. The transformer architecture outperforms CNN-based models across all noise levels, and query durations. In low noise conditions it achieves 47.9% for 1-sec queries, and 97% for 10-sec queries in finding the correct song, surpassing by 14%, and by 18,5% the second-best performing model, respectively. Under heavy noise levels, we achieve a detection rate 56,5% for 15-second query duration. All Experiments are conducted on a public large-scale dataset of over 100K songs, with queries matched against a database of 56 million vectors. We make our code publicly available on GitHub (https://github.com/magcil/deep-audio-fingerprinting-benchmark), supporting reproducibility and allowing future researchers to freely use our evaluation protocol to develop robust recognition systems for practical scenarios.

## 1    Introduction

Song identification is one of the oldest and perhaps one of the most popular tasks in Music Information Retrieval (MIR). It has attracted significant attention in both academic research and industry applications [1]. The goal of a song identification system is straightforward: suppose you hear a song in a café, restaurant, or any public space that you wish to learn more about it. Then, you capture the audio using a portable device. An algorithm processes the raw audio and extracts a compact representation from each predetermined and equally-length segment. In the literature, these representations are commonly referred to as audio fingerprints [2]. These fingerprints are then matched against a database of stored fingerprints, and the most similar entries are retrieved. A second data structure maintains the mapping between the fingerprints in the database and song metadata, providing all relevant information about the top-matching track. **Figure 1** provides a high-level overview this procedure.

To build an efficient and accurate music recognition system, a number of challenges should be addressed. First and foremost, the audio that is captured from the portable device is often degraded by background noise and other distortions that compromise signal quality. This significantly reduces the performance of the system as the query fingerprints may not match the fingerprints of the correct track in the database. Furthermore, as the number of songs in the database increases, the number of stored fingerprints increases as well. For these reasons, the fingerprints should be as compact as possible to lower the memory footprint, while maintaining their discriminative characteristics to be correctly matched among millions of other fingerprints. Another requirement is that the fingerprints should be easily computable to allow fast retrieval, capable of returning the correct answer in a few seconds to meet the needs of a real-world application. A thorough study of these challenges is presented in detail in [3].

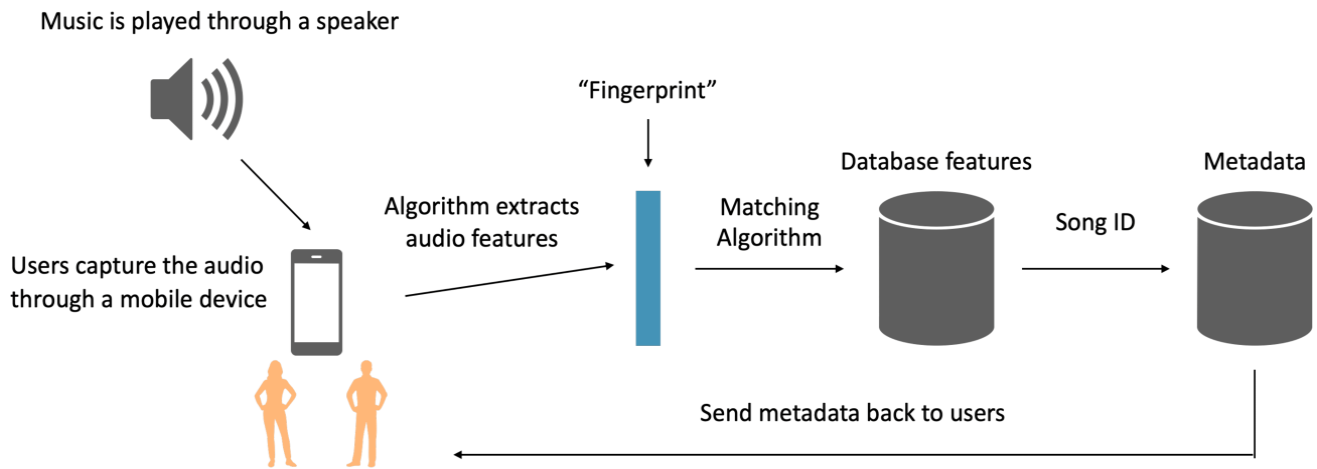

**Fig. 1.** A high-level overview of a music recognition system.

To address these challenges, Wang presented in [4] the pioneering algorithm that is used by Shazam[1], one of the most established music recognition platforms in use today. The algorithm focuses on identifying local maxima in the spectrograms of audio fragments. The key insight is that these peaks tend to be preserved even when the signal is heavily distorted or overlaid with other sound sources. **Figure 2** illustrates this idea: although the spectrogram of the distorted signal differs significantly from that of the original audio fragment, most of the spectral peaks are retained, as shown in the two images at the bottom. Using this observation, the correct match can be found by comparing the spectral peaks instead of the whole spectrograms. The audio fingerprint in this case is generated by hashing triplets of the form $(k_0, k_1, n_1 - n_0)$, where $(k_0, n_0),(k_1, n_1)$ correspond to neighbouring local maximum values in the spectrogram representation of the songs to be added in the database. A non-exhaustive technique based on inverted lists allows for fast and efficient retrievals. Several other methods [2, 5, 6, 7, 8] utilizing sophisticated techniques for audio fingerprinting have been developed, demonstrating notable performance. However, all these methods are inherently limited by their reliance on handcrafted features which leads to representations with significant storage requirements. Another drawback of these approaches is that they usually require long queries (>10 sec) to achieve high accuracy rates. This poses a limitation for on-device applications, where the memory footprint of the database should be as small as possible, and the response should be retrieved in a few seconds.

In contrast to these traditional methods based on handcrafted features, deep learning techniques have revolutionized numerous domains, including image classification [9, 10, 11] and object detection [12, 13, 14, 15], by learning feature representations directly from raw data. Motivated by these advancements, Google proposed an alternative approach for the song identification problem that leverages deep learning to derive compact, robust, and efficient audio representations, suitable for deployment on mobile devices [16]. In detail, they designed a Convolutional Neural Network (CNN) to map the 2D spectrogram representations of 2-sec audio fragments to a 96-dimensional feature vector. To ensure the robustness of these embeddings against signal distortions, the CNN was trained using triplet loss [17], minimizing the Euclidean distance between embeddings of original and distorted versions of the same song, while pushing apart embeddings from different songs. One key advantage of this approach is that the resulting feature representations can be efficiently compressed using product quantization, and retrieved using approximate nearest neighbour techniques [18, 19], significantly reducing memory usage and enabling fast search. In this setup, each song occupies on average less than 3KB in the indexed database. The system's performance was evaluated using short queries of audio fragments embedded with various types of background noise, over a dataset with a total duration of 450 hours. The system achieved a recall rate of 75.5% in correctly identifying the corresponding track. In [20], the authors extended the work of [16] by replacing the triplet loss with contrastive loss [21]. They scaled up the experiments using the Free Music Archive (FMA) dataset [22], consisting of 100K songs from various genres, spanning approximately 8,000 hours of audio. Additionally, they extended the prediction granularity from the song level to a more fine-grained audio search by restricting the prediction from the song-level to the segment-level prediction by allowing mismatches of less than $\pm 500$ms.

---

[1] https://www.shazam.com

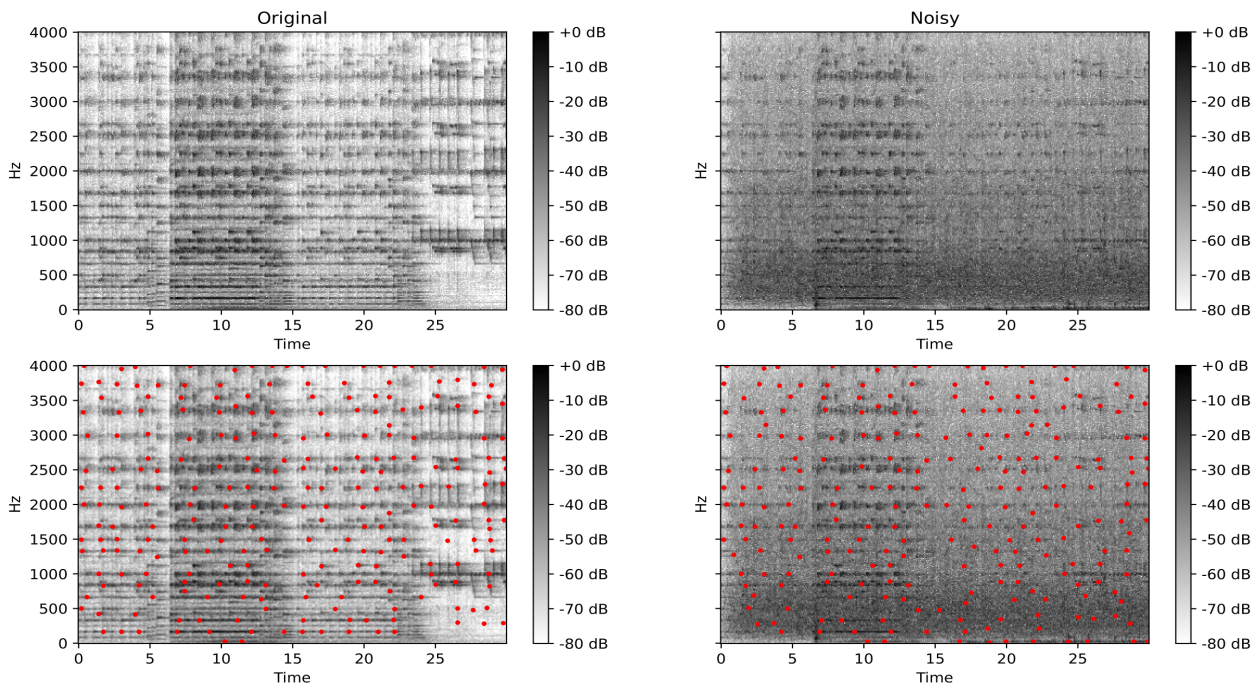

**Fig. 2.** Clean and distorted spectrogram corresponding to a 30-second audio fragment. While the distorted spectrogram differs significantly from its original counterpart, most of the spectral peaks have been retained.

The authors report performance improvements ranging from 7.8% for 10-second queries to 35.9% for 1-second queries, under an evaluation protocol in which the query audio fragments are distorted using the same augmentations applied during training with contrastive loss. In [23], an attention mechanism is integrated into a ResNet-based CNN backbone to further improve the system's robustness against distortions such as background noise and reverberation. The attention module is designed to mimic Wang's approach [4] by learning a mask during training that emphasizes local maxima in the feature maps of the CNN backbone. The authors use the FMA-Large dataset [22], consisting of 100K 30-second songs, to train the CNN with contrastive loss, and evaluate its performance on the FMA-Medium subset. They compare their approach to previous work [20] under evaluation scenarios where background noise with a fixed signal-to-noise ratio (SNR), ranging from 0dB to 15dB, is added to the test set. For 0 dB SNR, they report improvements of 6% for 5-second queries and up to 14.1% for 1-second queries. Several other works have extended the core ideas introduced in [16, 20] using deep learning methods, either to address the trade-off between database size and overall system performance [24, 25], or to accelerate the retrieval process [26]. However, a common limitation of existing work is the lack of consideration for system performance in practical scenarios where audio is captured using a mobile device's microphone in noisy and reverberant environments. To support this claim, we introduce an alternative evaluation protocol that more closely reflects real-world conditions. Using this protocol, we demonstrate that the performance of the systems presented in [20, 23] is significantly lower compared to the evaluation protocols used in these works. Our results suggest that these models, when evaluated under realistic conditions, do not generalize as well as previously claimed. This is due to the fact that in such cases more complicated signal degradations arise that are not reflected in the existing evaluation protocols. Furthermore, our evaluation framework allows us to identify specific weaknesses in these approaches and to propose methods for addressing them, ultimately improving performance and making these systems more suitable for real-world applications. In detail, our contributions in this work are summarized as follows:

1. We propose a novel evaluation protocol to assess the performance of music recognition systems under realistic conditions. Specifically, our protocol simulates scenarios where audio is captured using a mobile device in noisy and reverberant environments. We argue that systems performing well under this protocol more accurately reflect their effectiveness in real-world applications.

2. We demonstrate the effectiveness of our proposed protocol by reproducing two state-of-the-art (SOTA) deep learning-based systems [20, 23] under identical configurations. Our experiments reveal a significant performance drop when these systems are evaluated using our protocol compared to the original evaluation settings used in the respective works.

3. Our evaluation framework allows us to identify key weaknesses in existing systems and implement strategies to enhance their robustness. Specifically, we emphasize the critical role of data augmentation during training. By incorporating a simple transformation based on low/high pass filters we substantially increase performance across all models. Additionally, we apply transfer learning from a semantically related domain by introducing a lightweight projection module into a state-of-the-art transformer-based architecture. This results in a robust model that significantly outperforms previous approaches under our proposed evaluation protocol, making it ideal for practical scenarios.

## 2 Methodology

In this section, we provide an overview of the standard approach used to build deep learning-based music recognition systems. Specifically, we describe: (1) the training procedure for learning compact and discriminative audio embeddings using contrastive learning; (2) our proposed augmentation pipeline, which significantly improves model performance under our evaluation protocol and plays a critical role in counteracting real-world audio distortions; and (3) the database indexing strategy based on approximate nearest neighbor search with product quantization. This methodology has been widely adopted in prior work [16, 20, 23], typically combining metric learning—either contrastive or triplet loss—with efficient indexing techniques to reduce memory usage while enabling fast retrieval.

### 2.1 Contrastive Learning Framework

Contrastive learning is a powerful self-supervised approach for learning discriminative representations from raw data and has demonstrated state-of-the-art performance in domains such as image classification [21]. In the context of music recognition, it can be used to learn audio representations that are invariant to signal distortions. In detail, let $x \in \mathbb{R}^{F \times T}$ denote the spectrogram representation of an audio fragment, where $F$ is the number of frequency bins and $T$ is the number of time frames. Let $\mathcal{M}: \mathbb{R}^{F \times T} \rightarrow \mathbb{R}^{F \times T}$ be an augmentation function[2], that simulates distortions; thus, $\mathcal{M}(x)$ is the distorted version of $x$. Let $f_\vartheta: \mathbb{R}^{F \times T} \rightarrow \mathbb{R}^D$ denote the encoder network that maps spectrograms to D-dimensional embeddings, parameterized by $\vartheta$. The goal of contrastive learning is to pull together embeddings of positive pairs—such as $f_\vartheta(x)$ and $f_\vartheta(\mathcal{M}(x))$—while pushing apart embeddings of different songs. This is achieved using the contrastive loss function. Given a batch of clean spectrograms $\{x_1, \dots, x_N\}$ and their corresponding augmented versions $\{x_{N+1}, \dots, x_{2N}\}$ such that $x_{i+N} = \mathcal{M}(x_i)$, $i = 1, \dots, N$, the similarity-based loss between a positive pair $(x_i, x_j)$ is defined as:

$$\ell(i,j) = -log\left(\frac{\exp\left(<f_\vartheta(x_i),f_\vartheta(x_j)>\cdot T^{-1}\right)}{\sum_{k=1}^{2N} 1_{k \neq i} \cdot \exp(<f_\vartheta(x_i),f_\vartheta(x_k)>\cdot T^{-1})}\right) \quad (1)$$

where $<\cdot,\cdot>$ denotes the inner product, 1 the indicator function, and $T$ is a hyperparameter called temperature. Note that $\ell(i, j) \neq \ell(j, i)$ in general. The total contrastive loss over the batch is then computed as:

$$\mathcal{L} = \frac{1}{2N}\sum_{k=1}^{N}[\ell(k, N+k) + \ell(N+k, k)] \quad (2)$$

Algorithm 1 summarizes the training procedure with contrastive loss.

---

[2] Most augmentations are applied at the waveform level. For simplicity, we use spectrogram-level notation.

| | |
|---|---|
| **Algorithm 1: Training with Contrastive Loss** | |
| 1. | **Input:** *Encoder $f_\vartheta$, transformations $\mathcal{M}$.* |
| 2. | **Hyperparameters:** *Batch size N, temperature T>0, duration D>0.* |
| 3. | *Sample $\{y_1, \ldots, y_N\}$ audio fragments of duration D>0 from N different songs.* |
| 4. | *Compute the corresponding spectrograms $\{x_1, \ldots, x_N\}$.* |
| 5. | *For $k = 1, \ldots, N$ do* |
| 6. | *Compute the transformed samples $x_{N+k} = \mathcal{M}(x_k)$.* |
| 7. | *Set $z_k \leftarrow f(x_k)$.* |
| 8. | *Set $z_{N+k} \leftarrow f(x_{N/2+k})$.* |
| 9. | **End for** |
| 10. | *For $k = 1, \ldots, N$ do* |
| 11. | *Compute $\ell(k, k+N), \ell(N+k, k)$ according to (1).* |
| 12. | **End for** |
| 13. | *Compute $\mathcal{L} = \frac{1}{2N}\sum_{k=1}^{N}[\ell(k, k+N) + \ell(N+k, k)]$.* |
| 14. | *Update the parameters $\vartheta$ to minimize $\mathcal{L}$.* |

## 2.2 Augmentation Pipeline

The overall performance of the system is highly dependent on the choice of transformation $\mathcal{M}$ which aims to simulate signal distortions typically encountered in realistic conditions, such as background noise and reverberation. We adopt the augmentation pipeline from [20] referred to as the *baseline* augmentation. Through our proposed evaluation protocol, we find that this augmentation pipeline does not fully reflect all signal distortions encountered when audio is captured through a mobile device's microphone. For instance, as illustrated in **Figure 3,** low and high frequency components are often attenuated when audio is recorded through a mobile device's microphone—an effect not adequately addressed by the baseline transformations.

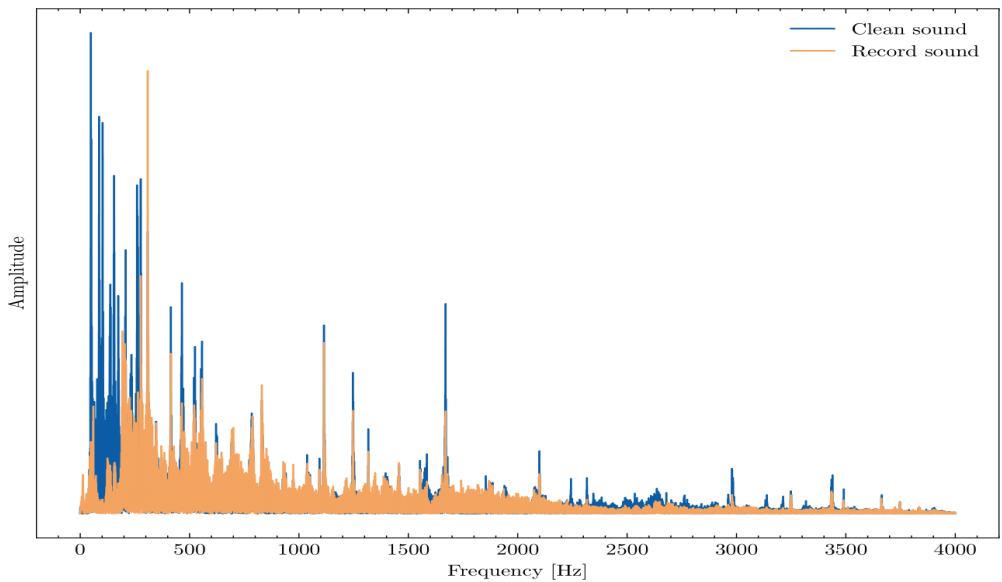

**Fig. 3.** Spectrum of the clean audio vs the recorded. Low and high frequencies tend to attenuate when the signal is recorded through the mobile device's microphone.

This is just one example of the many distortions that affect signal quality. To address this limitation, we augment the baseline pipeline with additional low-pass and high-pass filters. Specifically, we apply these filters to each audio fragment with a probability of 40%. For the low-pass filter, we randomly select a cutoff frequency $v_c \in (2000, 3000)$ Hz and a roll-off factor $\rho_c \in \{12, 24, 36\}$ dB. Frequencies in the range $(2^k v_c, 2^{k+1} v_c)$ for $k \in \mathbb{Z}_0^+$—i.e., the upper octaves of $v_c$—are attenuated by $k\rho_c$ decibels. Similarly, for the high-pass filter, we randomly sample a cutoff frequency $v_c \in (500, 1000)$ Hz and a roll-off factor $\rho_c \in \{12, 24, 36\}$ dB. In this case, frequencies in the range $(2^{k-1} v_c, 2^k v_c)$ for $k \in \mathbb{Z}_0^-$, are attenuated by $|k|\rho_c$ decibels. Denoting by $\mathcal{M}^B$ the baseline transformation and by $T_v$ the filtering transformation, our final augmentation pipeline is given by the composition $\mathcal{M}^* = T_v \circ \mathcal{M}^B$. Incorporating this extended transformation significantly improves the performance of both architectures in our proposed evaluation protocol.

## 2.3 Database Indexing

To index the database, we adopt the product quantization (PQ) method introduced in [27]. This approach is also employed in [20] and we apply it consistently across all experiments to ensure a fair comparison between different model architectures. Let $D$ be the dimension of the embedding space and $m$ a divisor of $D$. Product quantization partitions each embedding vector $x \in \mathbb{R}^D$ into $m$ equally length sub-vectors of dimension $D/m$. For each subspace, the method learns a codebook of $k^* = 2^k$ centroids via $K$-means clustering. Instead of storing the full-precision vectors, PQ encodes each subvector by the index of its closest centroid, effectively compressing the representation. The set of $m \cdot k^*$ centroids is stored in memory, while each original vector is approximated using a compact code of $m \cdot \log_2 k^*$ bits, significantly reducing the memory footprint compared to storing the original $32D$-bit float representations. To enable fast retrieval, a coarse quantizer is learned before applying product quantization. This quantizer assigns each database vector to one of a fixed number of clusters $K$, allowing the search to be restricted to a small subset of relevant clusters—significantly reducing the number of distance computations during query time. We follow a similar approach to previous works by extracting a feature vector from each 1-second audio segment with a 50% overlap. Under this configuration, a 3-minute song is represented by 359 encoded integer vectors.

## 3  Proposed Evaluation

To simulate a practical scenario and capture all particularities arising in such scenarios we propose the following protocol to effectively evaluate the performance of music recognition systems: We begin by normalizing 15 test songs to ensure equal energy levels and then concatenate them into a single continuous audio recording totalling 52 minutes in duration. In parallel, we generate a second 52-minute recording composed of various background noises. The clean song recording is played through a laptop, while a smartphone connected to a Bluetooth speaker simultaneously plays the noisy background recording. The resulting mixture is captured by the laptop's microphone. To account for spatial acoustic effects, the speaker is placed at three different distances from the laptop, resulting in three distinct recordings: *low.wav, mid.wav, and high.wav,* which reflect increasing levels of distortion. Specifically, *high.wav* corresponds to the closest speaker position (i.e., most distorted), while *low.wav* corresponds to the farthest (i.e., least distorted). System performance is evaluated using queries of varying durations from the set $\{1, 2, 3, 4, 5, 10, 15\}$ seconds. For each query segment, we retrieve the top-k nearest neighbours in the database and identify the predicted song through majority voting across all query segments. Accuracy is reported based solely on correct song-level matches. **Figure 4** demonstrates the procedure of generating the three recordings.

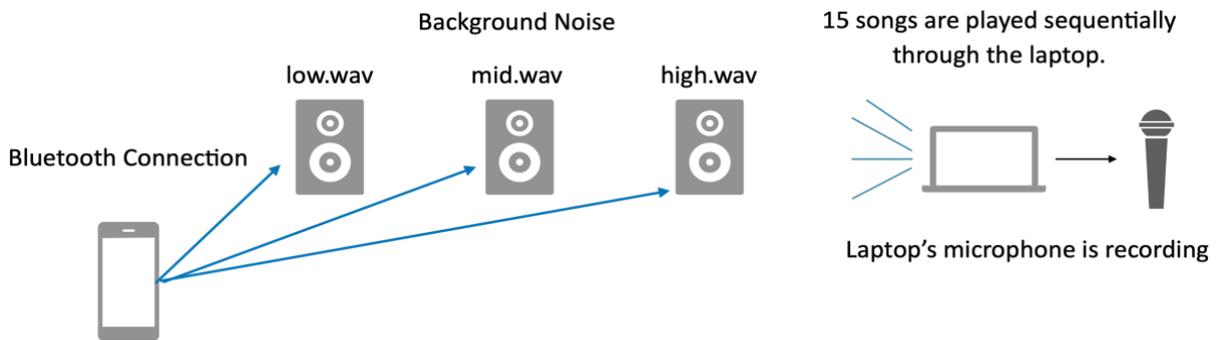

**Fig. 4.** Illustration of the proposed evaluation protocol.

## 4 Experimental Setup

### 4.1 Architectures

We build music recognition systems containing over 100K songs and approximately 56 million fingerprints, using the following three encoder architectures:

**Neural Audio Fingerprinter (NAFP)** is a CNN-based encoder $f: \mathbb{R}^{F \times T} \to \mathbb{R}^h$, composed of seven separable convolutional layers [28] which transform 2D spectrogram inputs into 1-dimensional feature vectors. An $\mathcal{L}_2$ normalization layer $g: \mathbb{R}^h \to \mathbb{R}^D$ is applied to project the encoder's output onto a D-dimensional unit sphere, yielding the final feature representation. We follow the exact configuration described in [20], where the input to the encoder is a mel-spectrogram with 256 mel-frequency bins and 32 time frames, corresponding to a 1-second audio segment. The output dimension of the encoder is $h = 1028$, and the projection dimension is $D = 128$. The normalization layer $g$ splits the encoder output into $D$ subvectors of size $h/D$, each of which is passed through a Linear–ELU–Linear block to produce a single scalar value. The dimension of the hidden layer before the projection to a single value is 32. The model has a total of 16,922,624 parameters.

**ResNet with Spatial-Temporal Attention (ResNet+STA)** is the architecture proposed in [23]. The encoder is based on a ResNet-like architecture [11], augmented with a spatial-temporal attention mechanism that emphasizes the local maxima of the extracted feature maps. To ensure compatibility with the NAFP model, we adapt the architecture to accept the same input spectrogram size, rather than the $64 \times 96$ input used in the original work. Additionally, we downsample the audio waveforms to 8kHz instead of 16kHz. This results in a significantly smaller dataset size—441 GB versus 882 GB—and ensures a fair comparison across all architectures under consistent settings. The model uses the same projection module as NAFP, producing a 128-dimensional unit-norm feature vector. In total, the architecture comprises 19,968,256 parameters.

**Bidirectional Encoder Representation from Audio Transformers (BEATs)** [29] is an iterative audio pretraining framework in which an acoustic tokenizer and an audio self-supervised model are optimized through alternating iterations. This approach has achieved state-of-the-art performance on sound event detection tasks. The BEATs model consists of 12 Transformer encoder layers [30], with 768 hidden units and 12 attention heads, totaling approximately 90 million parameters. The input features are 128-dimensional Mel filter banks, extracted using a 25ms Povey window and a 10ms frame shift. In contrast to the original implementation that uses 16kHz audio, we downsample the audio to 8kHz, resulting in input spectrograms of size $98 \times 128$. This configuration produces patch-level embeddings of shape $48 \times 768$. To obtain a single embedding vector, we average these patch-level embeddings and apply the same projection module used in the previous two architectures, producing a 128-dimensional unit-norm feature vector. We evaluate two configurations: training the model from scratch, and using transfer learning (**BEATs + TL**). For the transfer learning setup, we initialize the encoder with the weights of $\text{BEATs}_{iter3+}$, pretrained and fine-tuned on the AS2M dataset [31], and subsequently train the entire architecture using the same contrastive learning framework as in the other models.

### 4.2 Dataset

We use the full split of the FMA dataset [22] consisting of 106,574 full-length songs organized in a hierarchical taxonomy of 161 genres. We reserve 80% of the songs for training and 20% for validation. Additionally, we collect a total of 20 hours and 42 minutes of non-music background noise from the AudioSet dataset [31], which is split into training, validation, and test subsets in a 70-20-10 ratio. The test set is used to measure the performance of the systems on the baseline evaluation. To simulate reverberation effects for the augmentation pipeline, we gather 291 impulse response (IR) samples, including both public microphone responses [32] and spatial responses [33]. These are split 80-20 for training and validation, respectively. Finally, we download 15 commercial songs from YouTube using the yt-dlp package[3] to serve as the test set for our proposed evaluation protocol. These songs span a variety of genres (e.g., pop, rock, metal, classical), and are strictly excluded from the training phase.

---

[3] https://github.com/yt-dlp/yt-dlp

## 4.3 Implementation Details

Our codebase is implemented in Python, using PyTorch [34] as the deep learning framework. For audio analysis and feature extraction, we rely on the Librosa library [35], and we perform audio augmentations using the Audiomentations library[4]. We follow a similar approach as in [20]. Specifically, we filter out all 1-second audio segments with energy levels below 0 dB and use consistent training hyperparameters across all experiments. To construct a training batch of size $2N$, we first sample $N$ one-second segments from $N$ different songs. Each segment is then augmented using the transformation $\mathcal{M}$ producing a batch of $2N$ samples, comprising the original and their augmented versions. The feature extractor converts these audio samples into 2D spectro-temporal representations—mel-spectrograms for the CNN-based architectures and filterbanks for the Transformer-based model. The optimization procedure for each batch is outlined in Algorithm 1. Our training configuration uses a fixed sampling rate of $F_s = 8000$ Hz. All models are trained for 120 epochs with a batch size of $2N = 512$. We adopt the LAMB optimizer [36] with an initial learning rate of 0.003, decaying to $1 \cdot 10^{-7}$ using a cosine schedule. The embedding dimensionality is set to $D = 128$ in all experiments. To construct the fingerprint database, we extract one feature vector per 1-second audio segment using a sliding window with 50% overlap. We employ the Faiss library [37] to build an IVFPQ index, which stores quantized database vectors. The chosen index uses 32 sub-quantizers with 8-bit codebooks, resulting in each 128-dimensional vector being represented by a compressed code of $32 \cdot 8$ bits. The coarse quantizer for the inverted index contains 256 centroids, and at query time, we probe the 4 nearest clusters. All training is performed on a desktop PC equipped with an AMD Ryzen 9 7950X3D CPU and an NVIDIA GeForce RTX 4090 GPU. Training each model took approximately 5 hours. A summary of the hyperparameters used for database construction is provided in Table 1.

**Table 1.** Configuration Statistics.

|  |  |
|---|---|
| Number of Songs | 106,589 (441 GB) |
| Number of Vectors | 58,879,329 (30 GB) |
| Embedding Dimension (D) | 128 |
| Number of Sub-quantizers (m) | 32 |
| Codebook Bits | 8 |
| Coarse Quantizer K length | 252 |
| Faiss Index Size | 2,2 GB |
| N Probes | 4 |

## 5 Results and Discussion

### 5.1 Dejavu vs. deep-learning based fingerprinting

Before presenting the main experimental results on the large-scale dataset, we first compare a Shazam-like audio fingerprinting approach using the open-source library Dejavu [38] with the deep-learning approach on our evaluation protocol. This preliminary experiment is conducted on a smaller-scale dataset comprising a total of 26,016 songs. Specifically, 24,985 tracks are taken from the medium split of the FMA dataset, which consists of 30-second audio clips. An additional 1,031 full-length songs are included, bringing the total to 26,016 songs. Combined with the 15 test songs used in our proposed evaluation protocol, this setup yields a total of 58,659,291 fingerprints in the Dejavu database, occupying approximately 6.7 GB of storage.

We compare this system against the one based on the NAFP model, trained both with the baseline augmentation pipeline presented in [20] and with our proposed augmentation strategy discussed in Section 2.1. For indexing the database in the deep learning-based systems, we use an IVFPQ index with $m = 64$ subquantizers and a coarse quantizer with $K = 20$ centroids. The total number of fingerprints in this case is 1,973,384, and the resulting index occupies just 137 MB—substantially smaller than the Dejavu database. Table 2 presents the accuracy of each system in retrieving the correct match across the three audio recordings—*low.wav*, *mid.wav*, and *high.wav*—for varying query lengths: 2s, 5s, 10s, and 15s.

---

[4] https://github.com/iver56/audiomentations

Across all query durations and distortion levels, the NAFP system trained with the proposed augmentation strategy ("Ours") consistently achieves the highest accuracy—except at the longest query length (15 seconds), where Dejavu slightly outperforms it by a margin of 2.35%. By comparing the NAFP model trained with the baseline augmentations to the one trained with our proposed augmentation strategy, we observe a substantial improvement in performance across all query lengths and distortion levels. For instance, at a 2-second query with high distortion, our approach achieves an accuracy of 41.92% compared to 19.47% for the baseline. At a 5-second query, the proposed method surpasses the baseline by a significant margin of 31.42% on the most challenging (*high.wav*) recording. These results highlight the critical role of the augmentation pipeline in systems that extract audio fingerprints using deep neural networks trained with contrastive loss. Finally, we observe that Dejavu performs poorly for short queries ($\leq$ 5 seconds), while its performance increases substantially for longer queries ($\geq$ 10 seconds).

Table 2. Performance (Accuracy) of Dejavu vs Deep-Learning Based Systems on Proposed Evaluation.

| Method | Augmentations | Query Length (s) | low.wav | mid.wav | high.wav |
|---|---|---|---|---|---|
| NAFP | **Ours** | 2 | **85.03** | **62.65** | **41.92** |
| NAFP | Baseline | | 67.22 | 49.02 | 19.47 |
| Dejavu | - | | 12.63 | 6.50 | 10.73 |
| NAFP | **Ours** | 5 | **92.66** | **80.06** | **62.52** |
| NAFP | Baseline | | 80.38 | 65.39 | 31.10 |
| Dejavu | - | | 59.21 | 40.94 | 47.24 |
| NAFP | **Ours** | 10 | **94.52** | **90.32** | **74.19** |
| NAFP | Baseline | | 84.19 | 78.39 | 41.29 |
| Dejavu | - | | 83.60 | 70.98 | 71.29 |
| NAFP | **Ours** | 15 | **97.56** | **90.24** | 80.98 |
| NAFP | Baseline | | 86.34 | 80.98 | 48.29 |
| Dejavu | - | | 93.33 | 77.14 | **83.33** |

## 5.2 Large Scale Experiments

In this section, we present experiments conducted on the full FMA dataset, following the configuration outlined in Table 1. We begin by reporting results on the baseline evaluation protocol proposed in [23]. In this setup, we report the Top-1 hit rate, defined as:

$$100 \times \frac{(n \text{ of hits @ Top-1})}{(n \text{ of hits @ Top-1}) + (n \text{ of miss @ Top-1})}$$

where a match is considered correct if the retrieved timestamp of the query in the correct audio falls within a $\pm 500$ ms window. Note that this metric is upper-bounded by the accuracy of correctly identifying the matching audio clip. As shown in Table 3 all models perform reasonably well under this evaluation protocol, even for short query durations. For instance, with a query length of 2 seconds and an SNR of 0dB—representing one of the most challenging conditions—all models achieve at least around 44% accuracy, with the best model (BEATs+TL) reaching 58.5%. As the query length increases, the performance of all models improves steadily across all noise levels, indicating that longer queries provide more reliable temporal context for matching. Among the models, NAFP with our proposed augmentation pipeline consistently outperforms its baseline counterpart across all conditions. Notably, for a 5-second query at 5dB SNR, our method achieves 90.2% accuracy, a significant improvement over the 84.5% baseline. Additionally, the BEATs+TL model exhibits strong performance, often surpassing other architectures, especially for the shortest queries. This suggests that transformer-based architectures benefit from transfer learning and can generalize well to noisy, short queries when properly fine-tuned.

Table 3. Large Scale Experiment – Performance of different approaches on Baseline Evaluation.

| Model | Augmentations | Query [s] | 0 dB | 5 dB | 10 dB | 15 dB |
|---|---|---|---|---|---|---|
| NAFP | Baseline | | 34.1 | 58.3 | 72 | 80.4 |
| NAFP | Ours | | 36.2 | 59.1 | 73.3 | **81.9** |
| ResNet+STA | Baseline | 1 | 30.9 | 55.9 | 70.8 | 79.1 |
| ResNet+STA | Ours | | 29.5 | 54.7 | 70.2 | 79.1 |
| BEATs | Ours | | 27.4 | 50 | 65.1 | 76.3 |
| BEATs+TL | Ours | | **37.3** | **59.9** | **73.4** | 81.2 |
| NAFP | Baseline | | 52.8 | 76.3 | 84.9 | 90.4 |
| NAFP | Ours | | 57.2 | **79.5** | 88.2 | 92.3 |
| ResNet+STA | Baseline | 2 | 53.1 | 75.4 | 84.8 | 89.4 |
| ResNet+STA | Ours | | 48.7 | 73.8 | 83.9 | 89.9 |
| BEATs | Ours | | 44.2 | 71.3 | 82.9 | 89.2 |
| BEATs+TL | Ours | | **58.5** | 79.3 | **88.5** | **92.4** |
| NAFP | Baseline | | 62.9 | 81.2 | 88.2 | 92.5 |
| NAFP | Ours | | 65.5 | 83.4 | **91.5** | **94.2** |
| ResNet+STA | Baseline | 3 | 57.7 | 80.7 | 87.7 | 91.5 |
| ResNet+STA | Ours | | 57.7 | 80.7 | 88.1 | 91.7 |
| BEATs | Ours | | 56.7 | 79.5 | 88.5 | 93.4 |
| BEATs+TL | Ours | | **68.8** | **85.4** | 90.6 | 92.2 |
| NAFP | Baseline | | 67.3 | 83.9 | 89.5 | 93.3 |
| NAFP | Ours | | **71.8** | **88.6** | **91.9** | **94.6** |
| ResNet+STA | Baseline | 4 | 64.2 | 82.5 | 88.6 | 92.1 |
| ResNet+STA | Ours | | 64.1 | 82.6 | 88.3 | 92.6 |
| BEATs | Ours | | 61.6 | 81.8 | 88.6 | 92.7 |
| BEATs+TL | Ours | | 71.4 | 84.4 | 89.8 | 93.1 |
| NAFP | Baseline | | 72.2 | 84.5 | 89.7 | 92.4 |
| NAFP | Ours | | 74.4 | **90.2** | **91** | **94.2** |
| ResNet+STA | Baseline | 5 | 68 | 83.6 | 89.2 | 92.6 |
| ResNet+STA | Ours | | 67.3 | 86 | 90.8 | 92.9 |
| BEATs | Ours | | 65 | 85 | 90 | 93 |
| BEATs+TL | Ours | | **76.7** | 87.7 | 90.2 | 93.6 |

Although the results in Table 3 might be promising, they can also be misleading. This is because that this baseline evaluation it does not fully reflect the true performance of these systems in realistic use case scenarios—such as when audio is captured through a mobile device in a noisy and uncontrolled environment. This limitation becomes evident when deploying these models in real-world conditions, where their accuracy drops significantly compared to the controlled baseline results, where the background noise in added in an offline manner. This is also reflected from the results in our proposed evaluation in Table 4. In this case, we report the accuracy of the system; i.e., the number of times the system retrieved the correct song. To obtain the final result, we retrieve the 4 nearest neighbors for each 1-sec query segment. The winner is declared the song with the most occurrences. A comparison between the baseline evaluation (Table 3) and our proposed evaluation (Table 4) reveals a huge discrepancy in performance across all models. While models like NAFP and BEATs+TL achieve Top-1 hit rates exceeding 90% under the baseline protocol, their accuracy drops substantially in our proposed evaluation—especially for short queries and higher distortion (e.g., NAFP baseline drops to just 4.3% at 1s/*high.wav*).

Table 4. Large Scale Experiment - Performance of different approaches on the Proposed Evaluation.

| Model | Augmentations | Query [s] | low.wav | mid.wav | high.wav |
|---|---|---|---|---|---|
| NAFP | Baseline | 1 | 25.1 | 13.5 | 4.3 |
| NAFP | **Ours** | | 30.9 | 20.4 | 10.5 |
| ResNet+STA | Baseline | | 25.7 | 14.5 | 4.2 |
| ResNet+STA | **Ours** | | 27.7 | 16.4 | 6.2 |
| BEATs | **Ours** | | 21.7 | 11.4 | 4 |
| BEATs+TL | **Ours** | | **44.9** | **29.7** | **12.1** |
| NAFP | Baseline | 2 | 32.3 | 18 | 6 |
| NAFP | **Ours** | | 40.5 | 25.5 | 12.9 |
| ResNet+STA | Baseline | | 34.3 | 19.4 | 12.8 |
| ResNet+STA | **Ours** | | 36.7 | 24 | 8.1 |
| BEATs | **Ours** | | 28.7 | 14.5 | 5 |
| BEATs+TL | **Ours** | | **60.8** | **42.1** | **19.2** |
| NAFP | Baseline | 3 | 37.9 | 23.9 | 6.9 |
| NAFP | **Ours** | | 49.2 | 30.6 | 16.6 |
| ResNet+STA | Baseline | | 41.9 | 25.6 | 8 |
| ResNet+STA | **Ours** | | 44.8 | 29.3 | 10 |
| BEATs | **Ours** | | 36.9 | 21 | 6 |
| BEATs+TL | **Ours** | | **72.9** | **52.8** | **24.4** |
| NAFP | Baseline | 4 | 41.5 | 29.4 | 7.9 |
| NAFP | **Ours** | | 56.7 | 38.1 | 21.5 |
| ResNet+STA | Baseline | | 49.1 | 29.9 | 8.4 |
| ResNet+STA | **Ours** | | 52.2 | 35.7 | 13.2 |
| BEATs | **Ours** | | 42.9 | 25 | 9.4 |
| BEATs+TL | **Ours** | | **82.5** | **62.5** | **29.7** |
| NAFP | Baseline | 5 | 48.8 | 32.3 | 9 |
| NAFP | **Ours** | | 62.3 | 44.6 | 24.8 |
| ResNet+STA | Baseline | | 53.7 | 34.3 | 10 |
| ResNet+STA | **Ours** | | 58.6 | 41.3 | 14 |
| BEATs | **Ours** | | 48 | 30.6 | 19 |
| BEATs+TL | **Ours** | | **86.1** | **68.5** | **34.1** |
| NAFP | Baseline | 10 | 62.5 | 49 | 13 |
| NAFP | **Ours** | | 76.1 | 59.3 | 37.4 |
| ResNet+STA | Baseline | | 64.8 | 48.3 | 16.7 |
| ResNet+STA | **Ours** | | 71.6 | 56.7 | 20 |
| BEATs | **Ours** | | 64.8 | 43.5 | 15.4 |
| BEATs+TL | **Ours** | | **93.2** | **86.1** | **46.1** |
| NAFP | Baseline | 15 | 70.7 | 56 | 17.5 |
| NAFP | **Ours** | | 78.5 | 72.1 | 45.3 |
| ResNet+STA | Baseline | | 75.1 | 54.6 | 19.5 |
| ResNet+STA | **Ours** | | 76.5 | 67.3 | 28.7 |
| BEATs | **Ours** | | 70.7 | 53.1 | 17.5 |
| BEATs+TL | **Ours** | | **97** | **90.2** | **56.5** |

This discrepancy underscores the limitations of the baseline evaluation in capturing the complexities of real-world scenarios, such as noisy environments and mobile-recorded queries. Our proposed evaluation thus offers a realistic benchmark for system performance in these practical applications. Additionally, in this case, the augmentation transformations have a greater impact on the overall performance of the system. As it is evident, our proposed augmentation strategy consistently yields significant improvements compared to the baseline augmentations. For instance, NAFP with our augmentations improves over its baseline by up to 27.8% (at 15s/*high.wav*). Similar trends are observed across other query durations and distortion levels. A similar phenomenon is also observed for ResNET+STA model when trained with our proposed augmentations vs. its baseline counterpart.

BEATs+TL demonstrates the strongest performance under the proposed evaluation. It consistently outperforms all other models, achieving up to 97% accuracy at 15s/*low.wav*, maintaining a significant lead even under the most challenging conditions (e.g. at 15s/*high.wav*: 56.5%, compared to 45.3% for NAFP). In contrast, we observe that the transformer without transfer learning (BEATs) underperforms relative to the other models. In most cases, its performance is even lower than that of the CNN-based models without the proposed augmentations. This highlights the importance of transfer learning in this context. As the pretrained models obtained by BEATs have been trained in a self-supervised manner on a massive amount of audio data, they have already learned to extract discriminative audio representations. Additionally, an important factor of the performance boost obtained by transfer learning is that the AudioSet dataset is highly relevant to the music data in our case, as it contains many music-related excerpts. These observations underscore the fact that transformers require large-scale data to effectively learn the underlying distribution—transfer learning mitigates this need by initializing training with an already capable audio feature extractor. We also conducted experiments where the parameters of the encoder were frozen and only the projection module was trained. In this setting, the lightweight model underperformed, failing to learn effective representations. Additionally, we experimented with removing the projection module from the transformer-based architecture and using the average of the patch-level embeddings (of size $48 \times 768$) as the final 768-dimensional feature representation. This model also underperformed compared to **BEATs+TL**, highlighting the importance of the projection module in this architecture. An additional advantage of the projection module is its flexibility in controlling the embedding dimensionality. Unlike the fixed 768-dimensional embeddings, the projection module allows tuning to smaller sizes, which can significantly reduce storage requirements.

## 5.3   Performance – Quantization Trade-off

We conclude our experimental analysis by evaluating the impact of the quantization in the overall performance of the system. Assuming a 32-bit floating point representation, a 128-dimensional fingerprint possesses a total of $128 \cdot 32 = 4096$ bits in the database. This means, that the 58,879,329 unquantized fingerprints would require a total of 30 GB of space. In the case, of a product quantizer with $m$ quantizers and 8-bit codebooks, each 128-dimensional fingerprint is represented by a $8 \cdot m$ – bit integer vector of dimension $D/m$, therefore the code length in this case is $8 \cdot m$. There is a trade-off between the number of quantizers $m$ and the performance of the system. The smaller the number of quantizers $m$ the larger is the corresponding compression of the database, affecting the performance of the system. In this case, we vary the number of quantizers $m$ in the set $\{4, 8, 16, 32, 64, 128\}$. For each configuration, we report the accuracy achieved by the two best-performing models—NAFP and BEATs+TL—both trained with our proposed augmentation pipeline. Results are presented across the three recording conditions (*low.wav, mid.wav, high.wav*). The following figures illustrate the trade-off between retrieval accuracy and code length, along with the corresponding index size in megabytes. As shown, the optimal number of quantizers is $m = 32$, beyond which performance gains become marginal. In this configuration, the database size is approximately 2120 MB, a substantial reduction from the unquantized database size of 30 GB. In low-noise environments, (*low.wav*), BEATs+TL achieves relatively high performance (>70%) even with a code length of 128 (~ 1178 MB in index size). These experiments enable a more detailed analysis of the relationship between quantization and retrieval performance. As expected, stronger models allow for more aggressive quantization while still maintaining high accuracy.

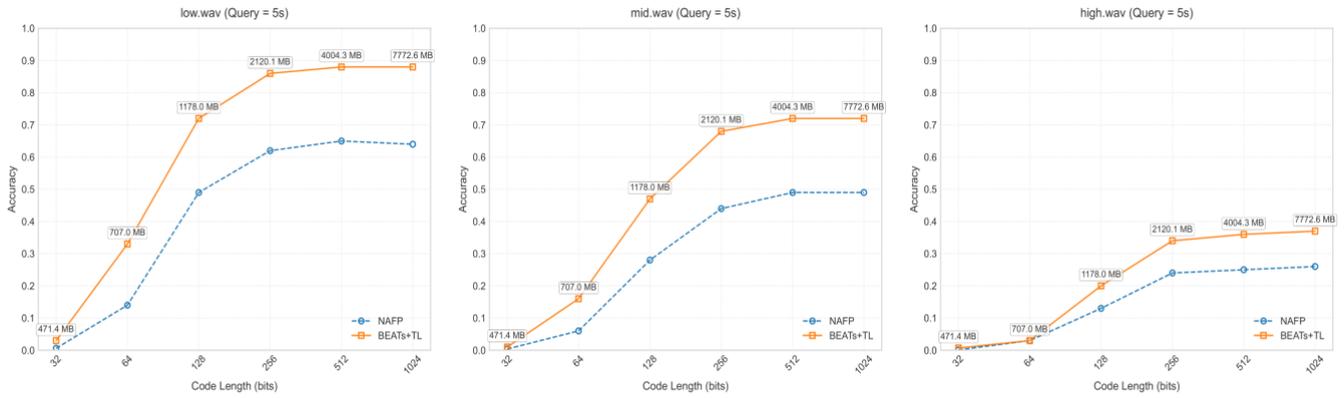

**Fig. 5. Accuracy vs. code length** for systems based on NAFP and BEATs+TL for query duration of 5 seconds.

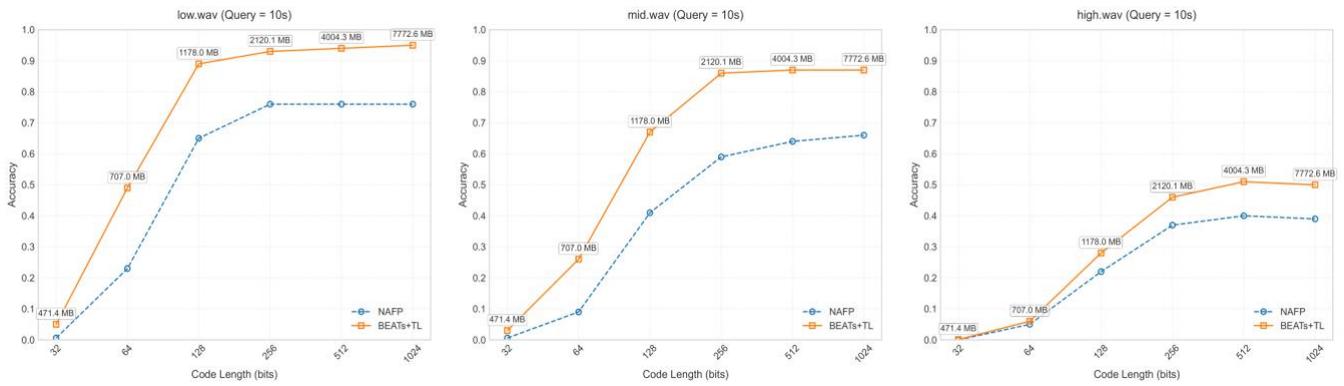

**Fig. 6. Accuracy vs. code length** for systems based on NAFP and BEATs+TL for query duration of 10 seconds.

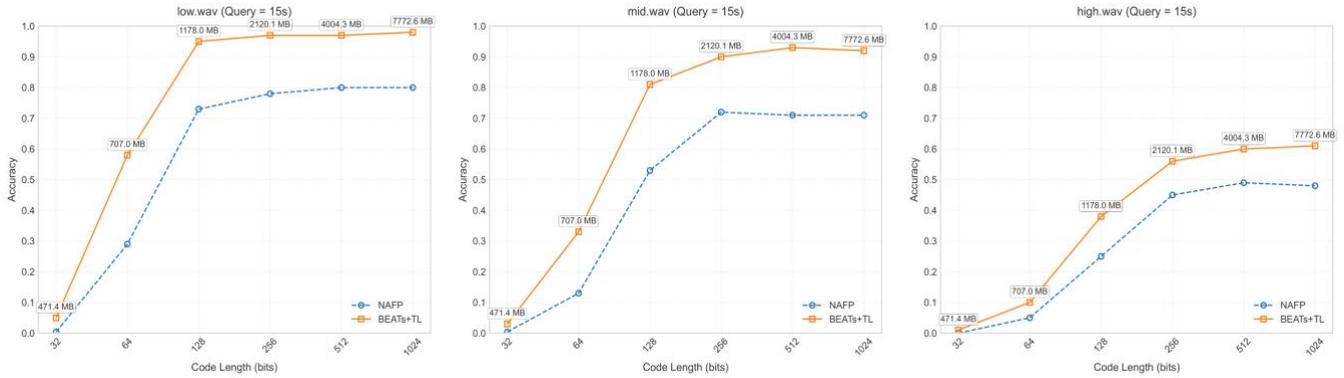

**Fig. 7. Accuracy vs. code length** for systems based on NAFP and BEATs+TL for query duration of 15 seconds.

## 6   Conclusions and Future Work

In this paper, we proposed a novel evaluation protocol to more accurately assess the performance of audio fingerprinting systems for song identification in realistic conditions. We conducted large-scale experiments using three deep learning architectures trained with contrastive learning and demonstrated that existing evaluation protocols fail to capture the challenges encountered in real-world scenarios—particularly when audio is captured through mobile devices in noisy environments. This was evidenced by the significant drop in accuracy under our proposed evaluation. To enhance system robustness, we refined the data augmentation process during contrastive learning by introducing low pass and high pass filters. Additionally, we showed that employing transfer learning in a transformer-based architecture, using a highly relevant source domain, further improved performance. Our transformer model, trained with the proposed augmentations, achieved state-of-the-art results under challenging conditions, making it highly suitable for practical scenarios. For future work, we aim to further investigate real-world distortions affecting signal quality and incorporate them into the training process to further increase model robustness. Moreover, we plan to integrate database indexing directly into the training phase. These end-to-end systems have been tested in the image domain [39] achieving promising results.

## Acknowledgments

The first author would like to express his gratitude to his colleagues from NCSRD, Eleanna Vali, Ioannis Koufos, and Marilena Sinni for reviewing the manuscript, providing valuable comments and observations.